\documentclass[usenatbib,twocolumn]{aastex7}

\usepackage{natbib}
\newcommand{\av}{\ensuremath{A(V)}}
\newcommand{\rv}{\ensuremath{R(V)}}
\newcommand{\ebv}{\ensuremath{E(B-V)}}
\newcommand{\nhi}{\ensuremath{N(HI)}}
\newcommand{\nhiebv}{\ensuremath{N(HI)/E(B-V)}}
\newcommand{\nhiav}{\ensuremath{N(HI)/A(V)}}
\newcommand{\fbump}{2175~\AA\ bump}
\newcommand{\alav}{\ensuremath{A(\lambda)/A(V)}}
\newcommand{\elvebv}{\ensuremath{E(\lambda - V)/E(B - V)}}

\shorttitle{M33 Ultraviolet Extinction}
\shortauthors{Gordon et al.}
\accepted{ApJ, 23 March 2026}

\begin{document}

\title{Large Variations Seen in First Ultraviolet Spectroscopic M33 Dust Extinction Curves}

\author[0000-0001-5340-6774]{Karl D. Gordon}
\affiliation{Space Telescope Science Institute, 3700 San Martin Drive, Baltimore, MD 21218, USA}
\affiliation{Sterrenkundig Observatorium, Universiteit Gent, Krijgslaan 281 S9, B-9000 Gent, Belgium}
\email[show]{kgordon@stsci.edu}

\author[0000-0002-9912-6046]{Petia Yanchulova Merica-Jones}
\affiliation{Institute of Astronomy and NAO, Bulgarian Academy of Sciences, 72 Tsarigradsko Chaussee Blvd., 1784 Sofia, Bulgaria}
\affiliation{University of Sofia, Faculty of Physics, 5 James Bourchier Blvd., 1164 Sofia, Bulgaria}
\affiliation{Space Telescope Science Institute, 3700 San Martin Drive, Baltimore, MD 21218, USA}
\email{petiay@gmail.com}

\author[0000-0002-0141-7436]{Geoffrey C. Clayton}
\affiliation{Space Science Institute,
4765 Walnut St., Suite B
Boulder, CO 80301, USA}
\email{
gclayton@spacescience.org}
\affiliation{Department of Physics \& Astronomy, Louisiana State University, Baton Rouge, LA 70803, USA}
\affiliation{Maria Mitchell Association, 4 Vestal St., Nantucket, MA 02554, USA}

\author[0000-0001-9806-0551]{Ralph Bohlin}
\affiliation{Space Telescope Science Institute, 3700 San Martin Drive, Baltimore, MD 21218, USA}
\email{bohlin@stsci.edu}

\author[0000-0001-9462-5543]{Marjorie Decleir}
\affiliation{European Space Agency (ESA), ESA Office, Space Telescope Science Institute, 3700 San Martin Drive, Baltimore, MD 21218, USA}
\altaffiliation{ESA Research Fellow}
\email{mdecleir@stsci.edu}

\author[0000-0002-7743-8129]{Claire E. Murray}
\affiliation{Space Telescope Science Institute, 3700 San Martin Drive, Baltimore, MD 21218, USA}
\affiliation{Department of Physics \& Astronomy, Johns Hopkins University, 3400 N. Charles Street, Baltimore, MD 21218}
\email{cmurray1@stsci.edu}

\author[0000-0001-9806-0551]{Luciana Bianchi}
\affiliation{Department of Physics \& Astronomy, The Johns Hopkins University, 3400 N. Charles St., Baltimore, MD 21218, USA}
\email{lbianch1@jhu.edu}

\begin{abstract}
Dust extinction curves provide one of the main avenues to understanding the detailed nature of dust grains and accounting for the effects of dust on observations of many astrophysical objects.
For the first time, spectroscopic ultraviolet (UV) extinction curves are measured in M33 expanding the sample of Local Group galaxies with such measurements to five.
These curves are based on Hubble Space Telescope/Space Telescope Imaging Spectrograph spectra and literature photometry from the UV to the near-infrared.
The four measured curves show large variations in their UV shapes including their 2175~\AA\ bump and UV slope strengths.
The average extinction of these four sightlines is lower than the averages for other Local Group Galaxies and does not follow the Milky Way $R(V)$ dependent relationship.
The variations between UV extinction shape parameters and gas-to-dust ratios for the M33 sightlines fall within the variations seen in the combined sample of UV extinction curves in the Milky Way, Large and Small Magellanic Clouds, and M31.
The correlation with gas-to-dust ratio is much stronger than the correlation with global metallicity.
This strengthens the picture that local conditions like radiation field density and shocks dominate over global galaxy properties like metallicity in determining the wavelength dependence of dust extinction.
\end{abstract}


\section{Introduction} \label{sec:intro}

Interstellar dust strongly affects the radiative transfer of photons in galaxies, is critical for molecular hydrogen formation and star formation, and modifies the observed spectral energy distributions (SEDs) of many astrophysical objects.
In the ultraviolet (UV) through infrared (IR), dust grains extinguish the SEDs of background sources through absorbing photons and scattering photons out of the observer's line-of-sight.
The wavelength dependence of this effect is an extinction curve.
Such extinction curves provide a wealth of information about dust grains themselves and allow for the effects of dust extinction to be accounted for in the observations of background objects.

UV extinction curves in the Milky Way (MW) have been extensively studied with samples of 400 or more showing significant variation in the overall UV slope and \fbump\ strength and width, but not in the centroid \citep{Witt84, Valencic04, Fitzpatrick07, Gordon09FUSE}.
MW studies have found that the UV through IR extinction curve can {\em on average} be parameterized with a single parameter chosen to be the total-to-selective extinction $\rv = \av/\ebv$ \citep{Cardelli89, Fitzpatrick19, Gordon23} and there are sightlines that significantly deviate from the \rv\ relationship \citep{Mathis92, Valencic03, Valencic04, Whittet04}.
Given our location inside the MW, UV extinction studies necessarily are dominated by the dust seen nearby within $\sim$1~kpc \citep{Valencic04}.

Measuring UV extinction curves in Local Group galaxies allows for the full census of variations to be studied from an external viewing geometry.
Measurements in the Large and Small Magellanic Clouds (LMC and SMC) have shown larger variations than those seen in the MW from a weakening of the \fbump\ strength near the 30~Dor star-forming region \citep{Clayton85, Fitzpatrick85, Misselt99} to the absence of the \fbump\ in most of the SMC \citep{Lequeux82, Prevot84, Gordon98, Gordon03, Gordon24}.
\citet{Gordon24} studied the combined behavior of the MW, LMC, and SMC extinction curves and found that there is a clear correlation in the UV extinction shapes with gas-to-dust ratio as measured by \nhiav.
This correlation is seen within the galaxies and the variations overlap significantly between galaxies.
The correlation with gas-to-dust is stronger than with metallicity indicating that dust grain growth/destruction through mass exchange with the gas phase is driving large variations in dust grain properties.

These correlations with \nhiav\ have been strengthened through recent UV extinction measurements in M31 \citep{Clayton25}.
The 17 M31 measurements were spread over a large portion of the M31 disk unlike, for example, observations in the Milky Way.
The behavior of the UV extinction parameters with \nhiav\ were seen to agree with the behavior seen for the MW, LMC, and SMC.
The best overlap was with the LMC in these correlations, again indicating that metallicity is not the dominant driver given M31 has a higher metallicity than the LMC.

This raises the question of what the UV extinction curves look like in M33, a Local Group galaxy at a similar distance as M31 with many UV bright OB stars.
M33 is a dwarf spiral galaxy with a metallicity $\sim$0.5 solar \citep{Bresolin11, Rogers22}.
The dust extinction in M33 has been studied from the UV to the IR using photometry from a variety of sources \citep{Wang22}.
In this work, the \av, \rv, and a coarse extinction wavelength dependence for over 125 sightlines towards OB supergiant stars was measured giving a range of \rv\ values from 2 to 6 and an average $\rv \sim 3.39$.
The coarse extinction curve revealed a weaker \fbump\ and slightly steeper far-UV rise than the MW average.
There also have been studies of attenuation curves for regions in M33 \citep{Gordon99, Hagen17, Williams19}.
While attenuation curves are more complicated to interpret as they include radiative transfer effects \citep{Witt96, Witt00}, they do provide insight into the extinction curves.
\citet{Gordon99} used radiative transfer models of the M33 nuclear stellar cluster to fit UV photometry and optical/NIR spectroscopy and found strong evidence for MW-like dust with a strong \fbump.
In contrast, \citet{Long02} obtained and analyzed UV spectroscopy of the nuclear cluster and did not see evidence for a strong \fbump.
The attenuation curves derived from a resolved attenuation analysis of M33 using SWIFT/UVOT imaging \citep{Hagen15, Hagen17} showed strong variations in the \rv\ and \fbump.
Their derived median UV attenuation curve has a weaker \fbump\ and is steeper than the \citet{Wang22} extinction curve.
This is in contrast with the radiative transfer analysis by \citet{Williams19} that found a quite strong \fbump\ and overall steep attenuation curve.
These works indicate that the M33 UV extinction is possibly different than the MW and shows significant variations, but confirmation is only possible with measured UV spectroscopic extinction curves.

This paper presents the first UV spectroscopic measurements of dust extinction in M33 using Hubble Space Telescope (HST)/Space Telescope Imaging Spectrograph (STIS) observations of OB supergiants.
Section~\ref{sec_obs} details the sample selection, HST/STIS observations and data reduction, and ancillary HST photometry.
The measurement of the extinction curves using stellar atmosphere models is given in Section~\ref{sec_ext}.
The variations within M33, average curve, and comparisons with other Local Group galaxies are discussed in Section~\ref{sec_discussion}.
Finally, Section~\ref{sec_summary} provides a summary.

\section{Observations}
\label{sec_obs}

\subsection{Sample}

\begin{deluxetable}{lllllllrl}
\label{tb_targets}
\tablecaption{Targets}
\tablehead{\colhead{Star\tablenotemark{a}} & \colhead{$\alpha_{2000.0}$} & \colhead{$\delta_{2000.0}$} & \colhead{V} & \colhead{SpT\tablenotemark{b}} & \colhead{UV SpT\tablenotemark{c}} & \colhead{ID} }
\startdata
J013250.80+303507.6 & 01 32 50.80 & +30 35 07.6 & 19.94 & B1:II: & B1 I & e1 \\
J013334.26+303327.6	& 01 33 34.26 & +30 33 27.6 & 18.55 & B2 I & B2 & e2 \\	 % PHATTER coords 01 33 34.228224 +30 33 27.39852
J013339.52+304540.5 & 01 33 39.52 & +30 45 40.5 & 17.50 & B0.5: I pec & O9 I & e3 \\   % PHATTER coords 01 33 39.481248 +30 45 40.50576
J013341.93+304728.3 & 01 33 41.93 & +30 47 28.3 & 18.52 & B1 I PCyg & B2.5 I & e4 \\   % PHATTER coords 01 33 41.884536 +30 47 28.31964
J013344.59+304436.9 & 01 33 44.59 & +30 44 36.9 & 19.79 & O If & B0 I & e5 \\   % PHATTER coords 01 33 44.636472 +30 44 36.66660		
J013406.63+304147.8	& 01 34 06.63 & +30 41 47.8 & 16.08 & O9.5 Ia & O9 I & e6 \\   % PHATTER coords 01 34 06.599904 +30 41 47.60088
J013410.59+304616.1 & 01 34 10.59	& +30 46 16.1 & 17.76 & B2.5 I & B2.5: I & e7 \\   % no PHATTER source, STIS data noise
J013416.10+303344.9 & 01 34 16.10 & +30 33 44.9 & 17.12 & B2.5 Ia  & B2.5 I & e8   % % PHATTER coords 01 34 16.071912 +30 33 44.74656
\enddata
\tablenotetext{a}{The star names are based on their celestial (J2000.0) positions \citep{Massey06}.}
\tablenotetext{b}{Spectral types from the literature \citep{Massey06, Humphreys14, Massey16}}
\tablenotetext{c}{Spectral types from STIS UV spectra (this work).}
\end{deluxetable}

\begin{figure}[tbp]
\epsscale{1.15}
\plotone{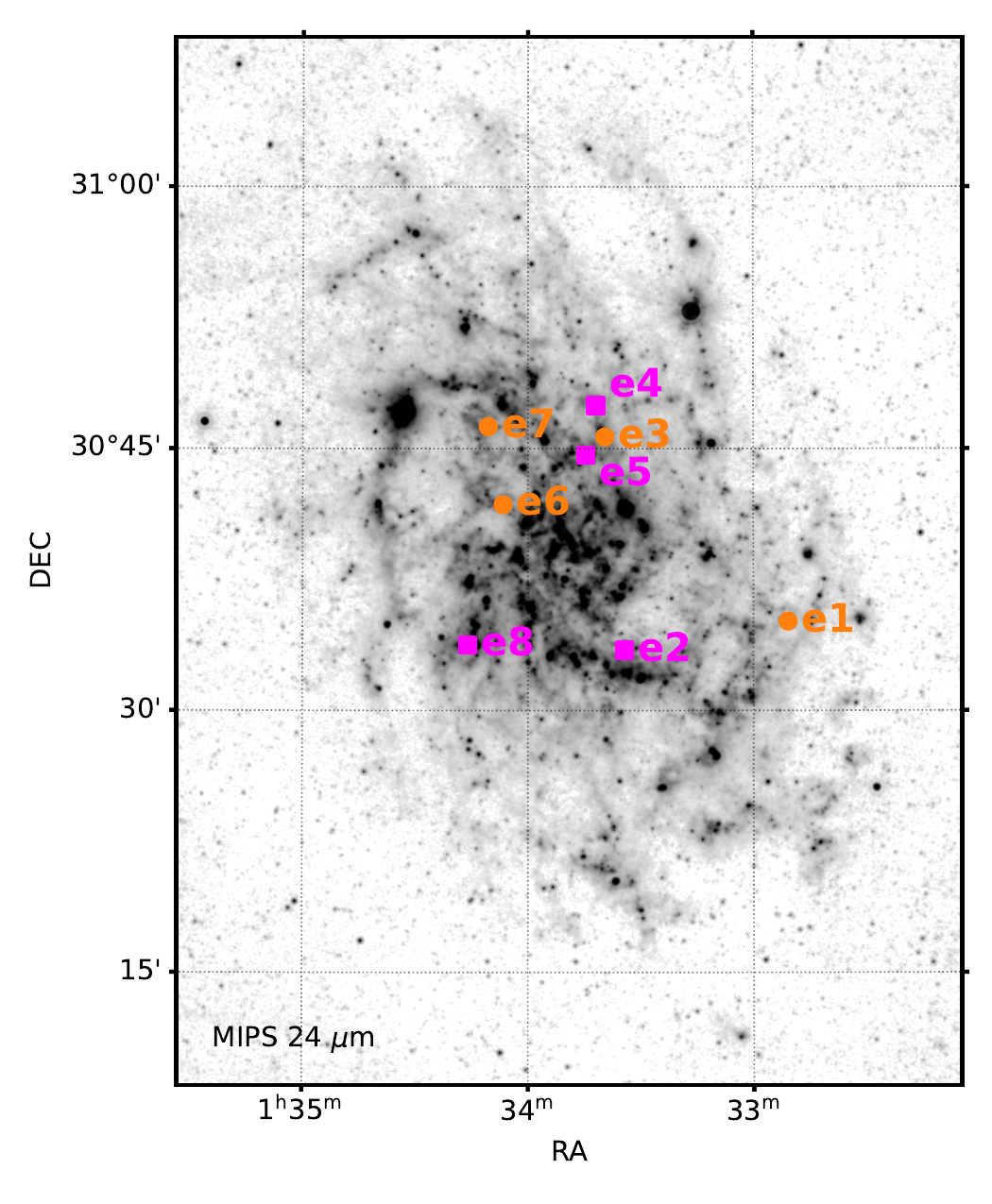}
\caption{The locations of the observed stars are shown on the Spitzer MIPS~24~\micron\ image \citep{Hinz04}.
This image traces the dust distribution with a focus on the youngest, embedded star formation.
The magenta squares give the sightlines where extinction curves were measured and the orange circles where this was not possible.
\label{fig_position}}
\end{figure}

The sample was selected from moderately reddened, early-type M33 stars which had ground-based spectral types \citep{Massey06, Humphreys14, Massey16}.
Then, HST PHATTER \citep{Williams21} images were examined to see if the star was single and well separated.
If the star turned out to be an OB association, then the PHATTER photometry was used to select an early-type moderately reddened star from the stars in the association.
Spectral typing using the STIS UV spectra was done using the grid provided in \citet{Neubig99}.
The stars and basic information about them are given in Table~\ref{tb_targets}.
The spatial distribution of stars superimposed on the Spitzer MIPS 24~\micron\ image \citep{Hinz04} is shown in Fig.~\ref{fig_position}.
This illustrates that the eight sightlines probe a large portion of the M33 disk.

\subsection{Data}
\label{sec_data}

HST/STIS low resolution spectroscopy was obtained using the G140L and G230L gratings for all 8 stars in our sample (PID: 15268).
The observations were taken with the $0\farcs 2 \times 0\farcs 2$ slit due to the crowded nature of the M33 regions targeted.
Each star was observed in 2 orbits, with an orbit for each grating setting.
Due to the faintness of the observations, the observations were reduced with custom steps as described by \citet{Clayton25}.
The spectra of e1 and e7 had such low signal-to-noise ratio that they were unsuitable for extinction curve measurements and were not used in this work.

Most of the stars were within the PHATTER footprint providing HST WFC3/ACS photometry from the UV to the NIR \citep{Williams21}.
For e3 the PHATTER photometry is missing for most possible bands and ground-based UBVRI photometry \citep{Massey06} is used instead.
The PHATTER and literature photometry is given in Table~\ref{tab_phot}.

\begin{deluxetable*}{lccccccc}
\label{tab_phot}
\tablecaption{Photometry}
\tablehead{\colhead{ID} & \colhead{WFC3/F275W} & \colhead{WFC3/F336W} & \colhead{ACS/F475W}	& \colhead{ACS/F814W} & \colhead{WFC3/F110W} & \colhead{WFC3/F160W} }
\startdata
e2 & $17.474 \pm 0.005$ & $17.505 \pm 0.004$ & $18.688 \pm 0.001$ & $18.377 \pm 0.001$ & $18.369 \pm 0.004$ & $18.230 \pm 0.002$ \\
e4 & $17.745 \pm 0.006$ & $17.629 \pm 0.005$ & $18.671 \pm 0.002$ & $18.186 \pm 0.002$ & $18.138 \pm 0.002$ & $17.992 \pm 0.002$ \\
e5 & $17.246 \pm 0.010$ & $17.519 \pm 0.003$ & $18.801 \pm 0.001$ & $18.404 \pm 0.001$ & $18.340 \pm 0.001$ & $18.118 \pm 0.002$ \\
e6 & $14.772 \pm 0.003$ & \nodata & \nodata & $15.650 \pm 0.003$ & $15.666 \pm 0.001$ & $15.389 \pm 0.001$ \\
e8 & $16.165 \pm 0.004$ & $16.084 \pm 0.003$ & $17.193 \pm 0.01$ & $16.887 \pm 0.010$ & $16.839 \pm 0.001$ & $16.678 \pm 0.001$ \\ \hline\hline
\colhead{ID} & \colhead{V} & \colhead{B-V} & \colhead{U-B} & \colhead{V-R} & \colhead{R-I} & \\ \hline
e3 & $17.503 \pm 0.004$ & $0.064 \pm 0.004$ & $-1.014 \pm 0.004$ & $0.099 \pm 0.004$ & $0.054 \pm 0.004$ \\
\enddata
\end{deluxetable*}

\begin{figure}[tbp]
\epsscale{1.2}
\plotone{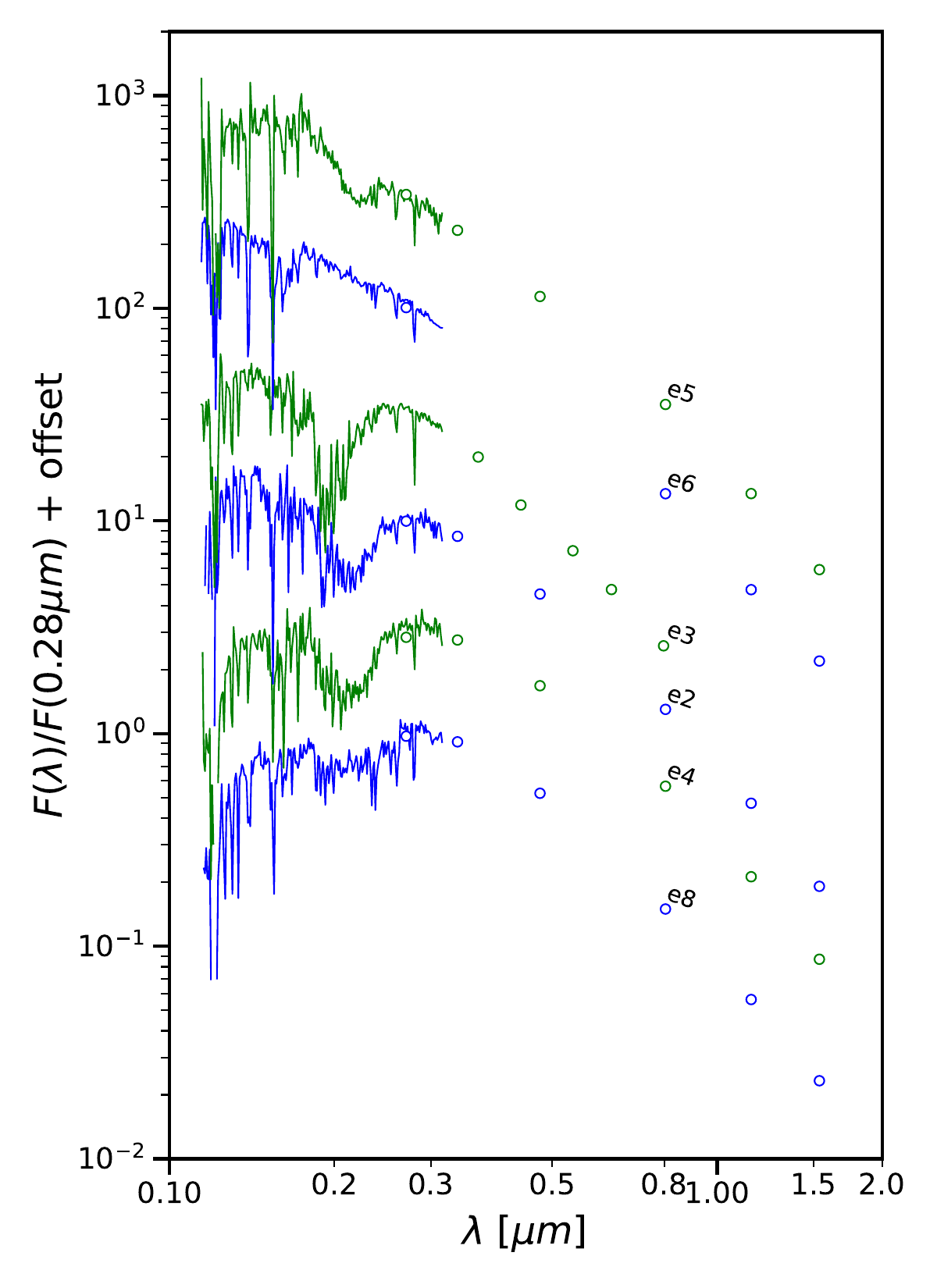}
\caption{The STIS spectra and photometry for the target stars are shown.
The spectra have been normalized to the flux at 2800~\AA, offset, and sorted by the UV spectral slope.
Regions of anomalously low flux have not been plotted.
\label{fig_spectra}}
\end{figure}

The UV spectra and UV/optical/NIR photometry are plotted in Fig.~\ref{fig_spectra}.
Visually examining the spectra, it is clear that the \fbump\ is seen in e2, e4, and e5.
It is less clear if the bump is seen in e3 given the strong stellar lines or instrumental issues shortward of the \fbump.
For e8, while there clearly is significant extinction given the overall spectral slope, the strength of the \fbump\ appears to be weaker than expected.
Finally, e6 does not show the presence of the \fbump, yet shows the same overall UV spectral slope as e5 that does show a \fbump.

\section{Extinction Curves}
\label{sec_ext}

The extinction curves were measured as was done for M31 \citep{Clayton25}.
In summary, the extinction curves were measured using the python package `measure\_extinction' \citep{measureextinction} by forward modeling the spectra and photometry based on stellar atmosphere models extinguished by foreground and internal dust and gas.
The stellar model atmospheres used were the recently updated non-LTE Tlusty models \citep{Lanz03, Hubeny25}.
The foreground extinction was modeled with a \citet[][aka G23]{Gordon23} $\rv = 3.1$ wavelength dependence with \av\ values based on the HI4Pi \citep{hi4pi16} radio measured \ion{H}{1} column densities integrated over MW velocities and the MW high-latitude measured \nhiebv\ ratio \citep{Liszt14}.
For the internal extinction, the UV was modeled with the FM90 functional form \citep{Fitzpatrick90}, the optical and NIR with the G23 \rv\ dependent model, and a cubic spline connecting these two segments \citep{Fitzpatrick19}.
We use a variant of the FM90 fitting where the \fbump\ amplitude is given by B$_3$ = C$_3$/$\gamma^2$ as B$_3$ directly measures the bump amplitude unlike C$_3$ \citep{Gordon24}.
The `dust\_extinction' package \citep{Gordon24JOSS} provided the FM90 and G23 models and the G23 model is based on literature extinction curve studies \citep{Gordon09FUSE, Fitzpatrick19, Gordon21, Decleir22}.
The MW foreground and M33 internal Ly$\alpha$ \ion{H}{1} absorption were modeled using Voigt profiles with the MW foreground component fixed to the HI4Pi measured value.

\begin{deluxetable}{clcccc}
\tablecaption{Model Parameters}
\tablehead{  & & & & & \colhead{Gaussian} \\
 \colhead{Parameter} & \colhead{Description} &
 \colhead{Units} &
    \colhead{Min} &
    \colhead{Max} &
    \colhead{(center, $\sigma$)}}
\startdata  
\multicolumn{5}{c}{M33 Component} \\ \hline
$\log(T_\mathrm{eff})$ & effective temperature & K & 4.18 & 4.74 & (x\tablenotemark{a}, 0.025)\\
$\log(g)$  & surface gravity & cm s$^{-1}$ & 1.75 & 4.75 & (y\tablenotemark{a}, 0.1) \\
$\log(Z)$  & metallicity & \nodata & -0.7 & 0.3 & (-0.3, 0.2) \\
$v_\mathrm{turb}$ & turbulent velocity & km s$^{-1}$ & 2 & 10 & \nodata \\
$A(V)$  & V band extinction & mag & 0.0 & 100.0 & \nodata \\
$R(V)$ & $A(V)/E(B-V)$ & \nodata & 1.5 & 7.0 & (3.0, 0.4) \\
$C_2$ & UV slope & & -0.1 & 5.0 & (0.73, 0.25) \\
$B_3$ & 2175~\AA\ bump height & \nodata & -1.0 & 8.0 & (3.6, 0.6) \\
$C_4$ & FUV curvature & \nodata & -0.5 & 1.5 & (0.4, 0.2) \\
$x_o$ & 2175~\AA\ bump centroid & $\micron^{-1}$ & 4.5 & 4.9 & (4.59, 0.2) \\
$\gamma$ & 2175~\AA\ bump width & $\micron^{-1}$ & 0.4 & 1.7 & (0.89, 0.08) \\ 
$\log(HI)$ & M33 \ion{H}{1} column & atoms cm$^{-2}$ & 16.0 & 24.0 & \nodata \\ 
$v(M33)$  & velocity & km s$^{-1}$ & \multicolumn{3}{c}{fixed, $-180$} \\ \hline
\multicolumn{5}{c}{MW Component} \\ \hline
$\log_\mathrm{MW}(H I)$ & MW H I column & atoms cm$^{-2}$ & \multicolumn{3}{c}{fixed, Table~\ref{tab_ext_col_param}} \\
$A(V)_\mathrm{MW}$ & MW dust column & mag & \multicolumn{3}{c}{fixed, Table~\ref{tab_ext_col_param}} \\
$R(V)_\mathrm{MW}$ & MW $A(V)/E(B-V)$ & \nodata &  \multicolumn{3}{c}{fixed, 3.1} \\
$v(MW)$  & velocity & km s$^{-1}$ & \multicolumn{3}{c}{fixed, 0} 
\enddata
\tablenotemark{a}{Gaussian centers x and y for $\log(T_\mathrm{eff})$ and $\log(g)$, respectively, were set by the UV spectral types given in Table~\ref{tb_targets}.}
\label{tab_fit_params}
\end{deluxetable}

As in \citet{Clayton25}, Bayesian fitting was used with all non-fixed parameters having either uniform priors on a bounded interval or truncated Gaussian priors based on Milky Way observations.
All the fitting parameters and prior information are given in Table~\ref{tab_fit_params}.
The main changes for the M33 extinction component from \citet{Clayton25} are a metallicity prior centered on $-0.3$ \citep{Rosolowky08}, a fixed M33 velocity of $-180$~km~s$^{-1}$, and expanding the \rv\ prior to be from 1.5 to 7.0.
Simultaneously fitting the stellar and dust parameters ensures that uncertainties in the stellar parameters are included in the extinction curve parameter uncertainties \citep{Fitzpatrick05}.
Ideally, the MW foreground dust parameters $\log_\mathrm(HI)$, $A(V)_\mathrm{MW}$, and $R(V)_\mathrm{MW}$ would be sampled during the Markov Chain Monte Carlo (MCMC) sampling based on their measured uncertainties.
This was attempted and the fits with emcee MCMC sampler \citep{emcee-Foreman-Mackey-2013} did not converge and were visibly worse than with fixed MW foreground parameters.
It is possible that using a different sampling technique that is better suited for large numbers of parameters (e.g., nested sampling) may provide better fits and this will be tested in future work.

\begin{figure*}[tbp]
\epsscale{1.2}
\plotone{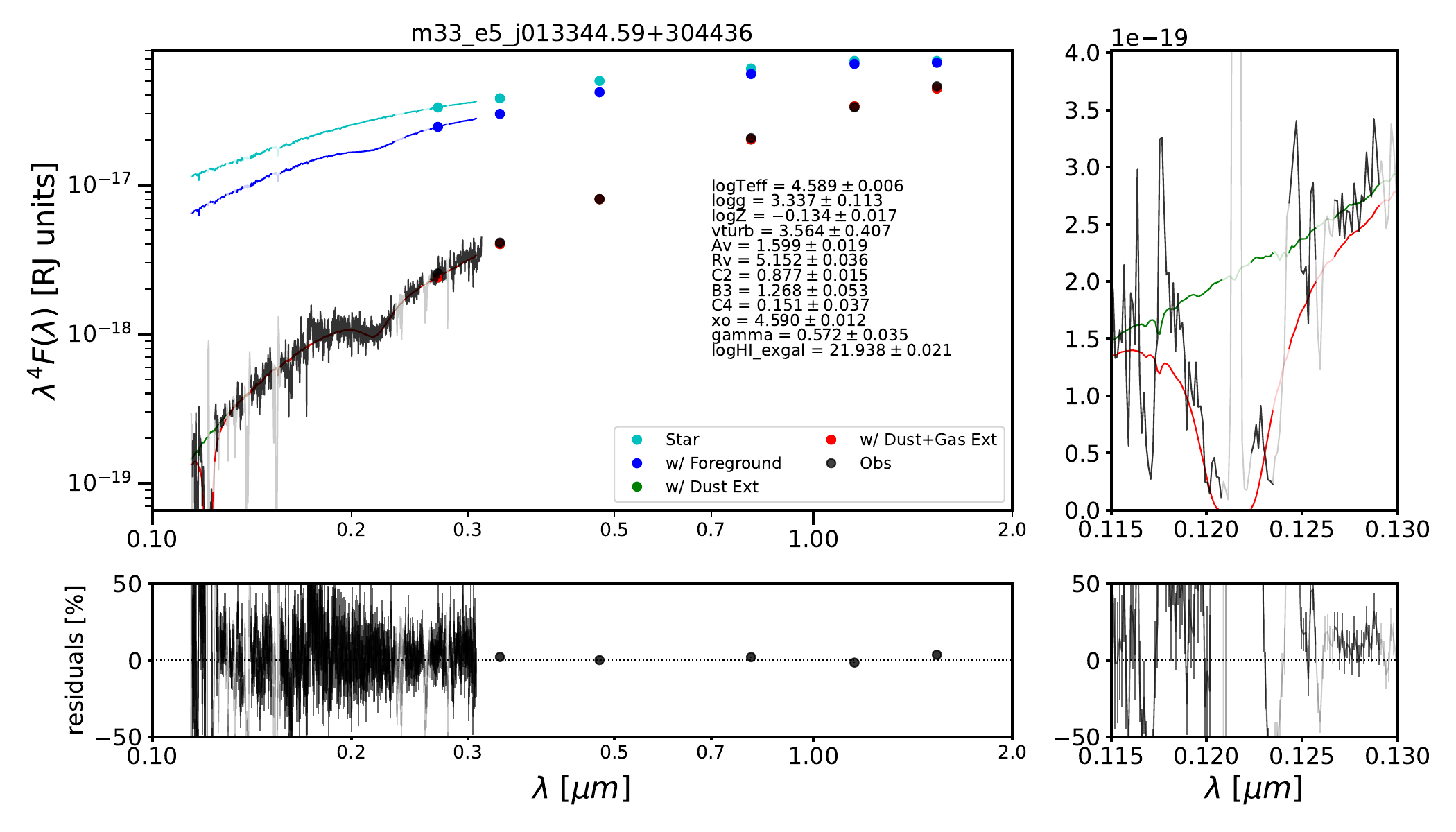}
\caption{The observations, model components, best fit model, and fit parameters are shown for e5.
In the left top panel, the full wavelength range is shown with the unextinguished model at the top (cyan), the MW foreground extinguished model in the middle (blue), and the MW foreground and M33 internal dust extinguished model (red) overplotted on the observations (black).
The right top panel gives the region around Ly$\alpha$ with the model with dust extinction only in green and the full model including the gas absorption in red.
The bottom panels give the residuals between the observations and the full model.
\label{fig_exp_fit}}
\end{figure*}

\begin{deluxetable}{lclcrc}
\tablecaption{Stellar Parameters \label{tab_ext_stell_params}}
\tablehead{\colhead{Name} & \colhead{$\log(T_\mathrm{eff})$} & \colhead{$T_\mathrm{eff}$} & \colhead{$\log(g)$} & \colhead{$\log(Z)$} & \colhead{$v_\mathrm{vturb}$} \\
 & \colhead{[K]} & \colhead{[K]} & \colhead{[cm s$^{-2}$]} & & \colhead{km s$^{-1}$}}
\startdata
e2 & $4.313 \pm 0.001$ & $20566 \pm 66$ & $2.305 \pm 0.005$ & $-0.499 \pm 0.001$ & $7.50 \pm 0.01$ \\
e4 & $4.306 \pm 0.001$ & $20240 \pm 38$ & $2.301 \pm 0.001$ & $-0.301 \pm 0.003$ & $5.00 \pm 0.02$ \\
e5 & $4.589 \pm 0.006$ & $38821 \pm 541$ & $3.337 \pm 0.113$ & $-0.134 \pm 0.017$ & $3.56 \pm 0.41$ \\
e8 & $4.227 \pm 0.001$ & $16861 \pm 51$ & $2.303 \pm 0.009$ & $0.170 \pm 0.016$ & $7.97 \pm 0.43$
\enddata
\end{deluxetable}

\begin{deluxetable}{lccccccc}
\tablecaption{Column Parameters \label{tab_ext_col_param}}
\tablehead{\colhead{Name} & \colhead{$A(V)$} & \colhead{$R(V)$} & \colhead{$log[N(HI)]$} & \colhead{$A(V)_\mathrm{MW}$} & \colhead{$log[N(HI)]_\mathrm{MW}$} \\
& & & \colhead{atoms cm$^{-2}$} & & \colhead{atoms cm$^{-2}$}}
\startdata
e2 & $0.938 \pm 0.008$ & $4.382 \pm 0.048$ & $21.64 \pm 0.05$ & $0.16$ & $20.64$ \\
e4 & $1.138 \pm 0.006$ & $3.768 \pm 0.038$ & $21.79 \pm 0.11$ & $0.16$ & $20.63$ \\
e5 & $1.599 \pm 0.019$ & $5.152 \pm 0.036$ & $21.94 \pm 0.02$ & $0.16$ & $20.63$ \\
e8 & $1.062 \pm 0.004$ & $6.994 \pm 0.008$ & $21.85 \pm 0.05$ & $0.16$ & $20.63$
\enddata
\end{deluxetable}

\begin{deluxetable}{lcccccc}
\tablecaption{FM90 Parameters \label{tab_ext_fm90_params}}
\tablehead{\colhead{Name} & \colhead{$C_2$} & \colhead{$B_3$} & \colhead{$C_4$} & \colhead{$x_o$ [$\micron^{-1}$]} & \colhead{$\gamma$ [$\micron^{-1}$]}}
\startdata
e2 & $1.32 \pm 0.03$ & $4.70 \pm 0.16$ & $-0.09 \pm 0.04$ & $4.726 \pm 0.013$ & $0.81 \pm 0.03$ \\
e4 & $1.48 \pm 0.02$ & $2.95 \pm 0.11$ & $0.04 \pm 0.07$ & $4.673 \pm 0.013$ & $0.74 \pm 0.03$ \\
e5 & $0.88 \pm 0.02$ & $1.27 \pm 0.05$ & $0.15 \pm 0.04$ & $4.590 \pm 0.012$ & $0.57 \pm 0.03$ \\
e8 & $3.01 \pm 0.03$ & $2.13 \pm 0.06$ & $0.19 \pm 0.07$ & $4.626 \pm 0.016$ & $1.19 \pm 0.04$ \\
Average & $1.46 \pm 0.03$ & $2.99 \pm 0.09$ & $0.43 \pm 0.05$ & $4.661 \pm 0.015$ & $1.41 \pm 0.08$ \\
\enddata
\end{deluxetable}

Fig.~\ref{fig_exp_fit} illustrates the extinction curve measurement technique showing the best fit model, components of the model, and the observations for e5.
Similar fits were obtained for e2, e4, and e8.
It was not possible to obtain a good fit of the e3 SED.
While this spectrum does show the turnover as expected for dust extinction with the UV being strongly depressed compared to the much less extinguished optical and NIR, it also shows very strong and sharp absorption features superimposed on the short wavelength side of the \fbump\ that is not expected (see Fig.~\ref{fig_spectra}).
Similar weaker sharp features were seen in UV spectra in M31 and these are attributed to artifacts due to the faintness of the targets and low sensitivity at these wavelengths \citep{Clayton25}.
In this case, the stronger sharp features make it challenging to measure the mid-UV extinction curve and hence an extinction curve was not measured for e3.
While it was possible to fit the e6 SED, the resulting extinction curve was quite flat in the UV without any detectable \fbump.
This star was studied in detail by \citet{Kourniotis18} where e6 is a binary star surrounded by a dense, circumbinary shell.
This binary is unsuitable for measuring an extinction curve and is not included in the M33 extinction curve sample as a result.
The e6 extinction curve was reminiscent of the two flat extinction curves seen in the SMC \citep{Gordon24}, which casts doubt that those curves are measuring interstellar dust.
The stellar, column density, and FM90 fit parameters for the four sightlines with good extinction curve measurements are given in Tables~\ref{tab_ext_stell_params}, \ref{tab_ext_col_param}, and \ref{tab_ext_fm90_params}.
The FM90 parameters are given in \elvebv\ units.

\begin{figure}
\epsscale{1.2}
\plotone{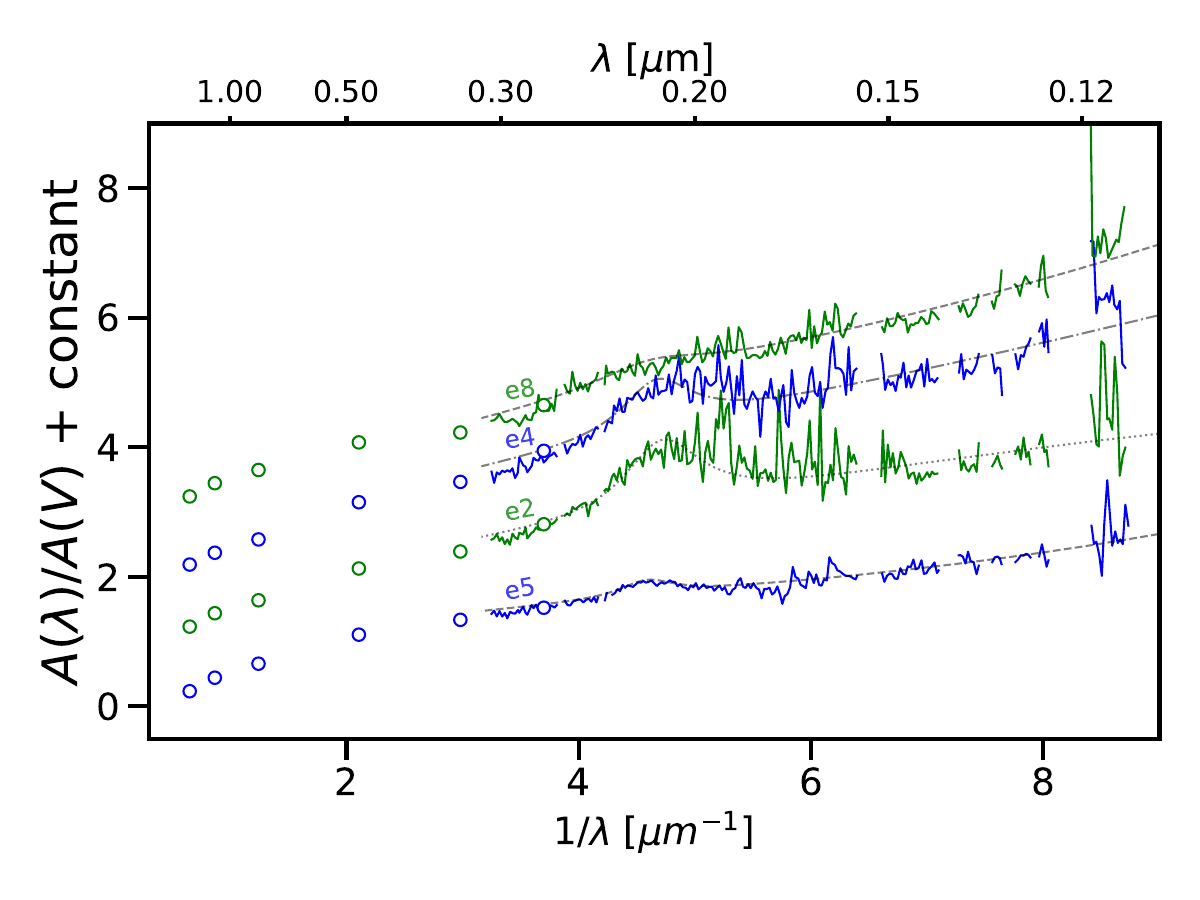}
\caption{The MW foreground-corrected M33 extinction curves are plotted sorted by UV slope in these units (i.e., $C_2 / R(V) + 1$).
The curves are rebinned to a resolution of 200.
The FM90 fits are plotted as non-solid lines.
For clarity, the curves are offset on the y-axis.
Regions of low S/N, near Ly$\alpha$, and around wind lines have been masked.
\label{fig_curves}}
\end{figure}

The extinction curves are calculated using model spectra based on the 50\% percentile posterior probability distribution function (pPDF) values for the stellar parameters and the assumed MW foreground extinction, which is equivalent to the standard pair method \citep{Massa83}; and including the MW foreground extinction in the ``comparison" means that the extinction curve only probes the dust in M33.
The extinction curves are measured relative to the $F475W$, $F814W$, or V bands and, where needed, are converted to the standard $V$ band using G23 extinction curves with \rv\ values as measured for each curve.
The \alav\ extinction curves for all four sightlines are shown in Fig.~\ref{fig_curves}.
The curves are mainly used for visualization, hence the use of 50\% pPDF values is reasonable.
It was deemed acceptable to use these curves to compute the average M33 curve (Sec.~\ref{sec_average}) as the small sample size and large variation between the curves dominates the uncertainty on the average, not the uncertainties on the individual curves.
Most of the following discussion focuses on the behavior of the FM90 parameters where the pPDF based uncertainties are used.

\section{Discussion}
\label{sec_discussion}

\subsection{Variations}

\begin{figure*}[tbp]
\epsscale{1.15}
\plotone{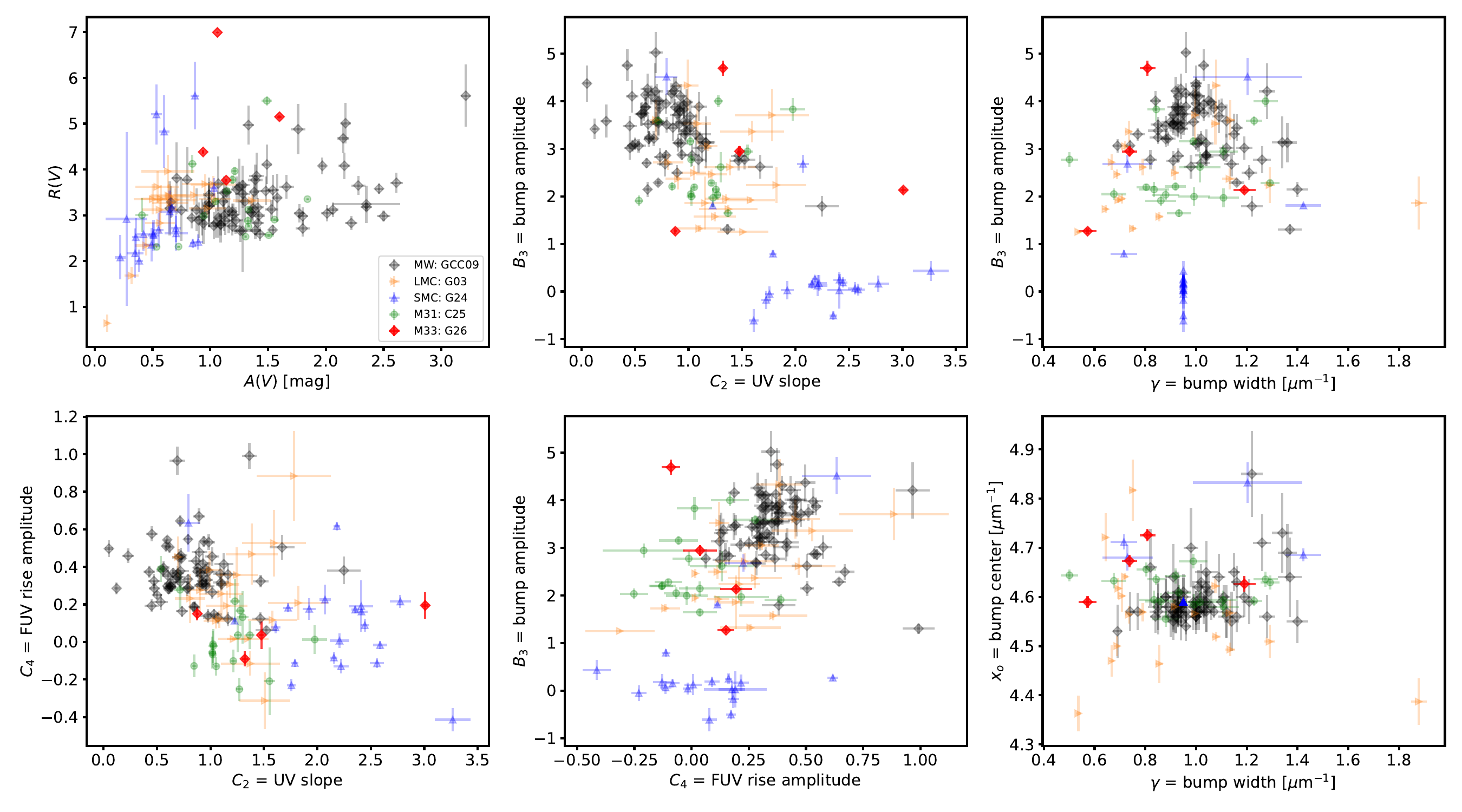}
\caption{\av\ versus \rv\ is shown in the upper left panel.
The other panels show different FM90 parameters versus each other.  
The $C_1$ versus $C_2$ is not shown as our fitting technique does not include $C_1$; instead $C_1$ is related to $C_2$ using the known strong correlations between these two parameters. 
The data sources are MW\_GCC09 \citep{Gordon09FUSE}, LMC\_G03 \citep{Gordon03}, SMC\_G24 \citep{Gordon24}, M31\_C25 \citep{Clayton25}, and this work (M33\_G25).
In the right two panels, the SMC blue points with single values of $\gamma$ and $x_o$ are for the sightlines without significant 2175~\AA\ bumps where these parameters were fixed in the fitting \citep{Gordon24}.
\label{fig_fm90}}
\end{figure*}

While only four extinction curves are presented, there is significant variation seen (Fig.~\ref{fig_curves}).
As the sightlines are spread throughout the M33 disk (Fig.~\ref{fig_position}) such large variations would likely be reflected in a larger sample.
Curves are seen that have significant, weak, and even non-existent 2175~\AA\ bumps and the UV slopes vary from mostly flat to fairly steep. 
Even with this small sample, the wavelength dependence of the dust extinction evidently varies significantly across M33.

The variations in the UV extinction curve shapes within M33 are compared to other Local Group galaxy measurements in Fig.~\ref{fig_fm90}.
The Local Group measurements consist of those from the Milky Way \citep{Gordon09FUSE}, LMC \citep{Gordon03}, the SMC \citep{Gordon24}, and M31 \citep{Clayton25}.
The upper left panel shows that the M33 values partially overlap the other galaxies, and the M33 \rv\ values are significantly higher on average.
Based on a larger sample of sightlines also towards OB stars, \citet{Wang22} found an average $\rv \sim 3.39$ indicating that our sample of four sightlines is likely biased to high \rv\ values.

The overall agreement between the different galaxies adds further evidence that there is a family of UV extinction curves that describes the behavior throughout the Local Group \citep{Gordon24, Clayton25}.
The M33 points clearly show that there is significant UV extinction curve shape variation in this galaxy and that this variation overlaps fairly well with the variations seen in the combined sample of the other four galaxies.
For example, the correlation between the \fbump\ strength ($B_3$) and the UV slope ($C_2$) indicate that as the \fbump\ weakens, the overall UV extinction slope strengthens.
Similar although weaker correlations are seen between the FUV rise amplitude ($C_4$) and the UV slope ($C_2$) and \fbump\ strength ($B_3$).
No obvious correlations are seen that involve the \fbump\ center ($x_o$) or width ($\gamma$).

\begin{figure*}[tbp]
\epsscale{1.15}
\plotone{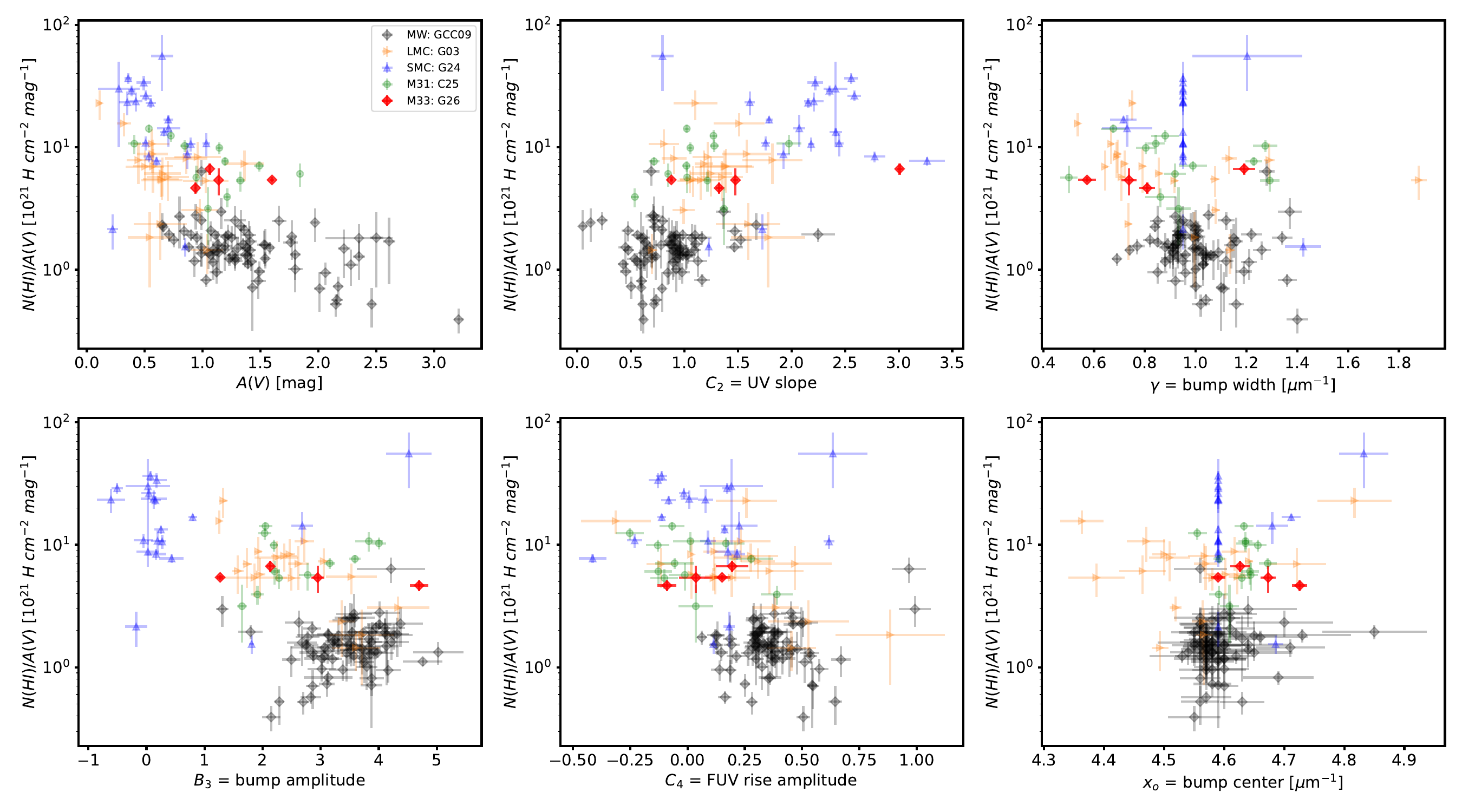}
\caption{The behavior of the FM90 extinction parameters versus \av\ and gas-to-dust ratio \nhiav\ are plotted.
The plot of \nhiav\ versus \av\ shows that generally the gas-to-dust ratio increases from the MW to the LMC to the SMC.
The data sources are MW\_GCC09 \citep{Gordon09FUSE}, LMC\_G03 \citep{Gordon03}, SMC\_G24 \citep{Gordon24}, M31\_C25 \citep{Clayton25}, and this work (M33\_G25).
In the right two panels, the SMC blue points with single values of $\gamma$ and $x_o$ are for the sightlines without significant 2175~\AA\ bumps where these parameters were fixed in the fitting \citep{Gordon24}.
\label{fig_gdprops}}
\end{figure*}

Fig.~\ref{fig_gdprops} plots \av\ and the FM90 parameters versus gas-to-dust ratio as measured by \nhiav\ for M33 and other Local Group galaxies.
\citet{Gordon24} found that the UV extinction curve shape variations correlated with \nhiav\ for extinction curves in the MW, LMC, and SMC.
This was strengthened by finding that M31 followed the same correlation \citep{Clayton25}.
The M33 points do not show much variation in gas-to-dust ratio, yet have significant variation in UV extinction shape parameters.
The distribution of M33 points does fit with the variations seen at similar gas-to-dust ratios in other galaxies indicating that the UV extinction shape parameters are not driven solely by gas-to-dust ratio.
This is not surprising as there is significant variation in extinction parameters in the MW points in these plots even though the range in MW gas-to-dust ratio and metallicity values is limited given they are measuring dust within $\sim$1~kpc of the Sun.
The MW variations are attributed to changes in \rv, indicating that \rv\ and gas-to-dust are both likely needed to explain the correlations seen in these plots.

The M33 points overlap with the M31 and LMC distributions more than the Milky Way or SMC distributions.
The similar gas-to-dust ratio values in M33 are in general agreement with the lack of metallicity variations in a dwarf galaxy like M33.
Yet, the large variations in gas-to-dust ratio in the LMC and SMC where the metallicities are also known not to vary significantly within each galaxy indicate that local conditions are important for dust extinction properties.
Given this and the range of gas-to-dust variation within each galaxy, the effects of local environment drive larger variations in extinction properties than global galaxy parameters like metallicity.

\subsection{Average}
\label{sec_average}

\begin{figure}[tbp]
\epsscale{1.15}
\plotone{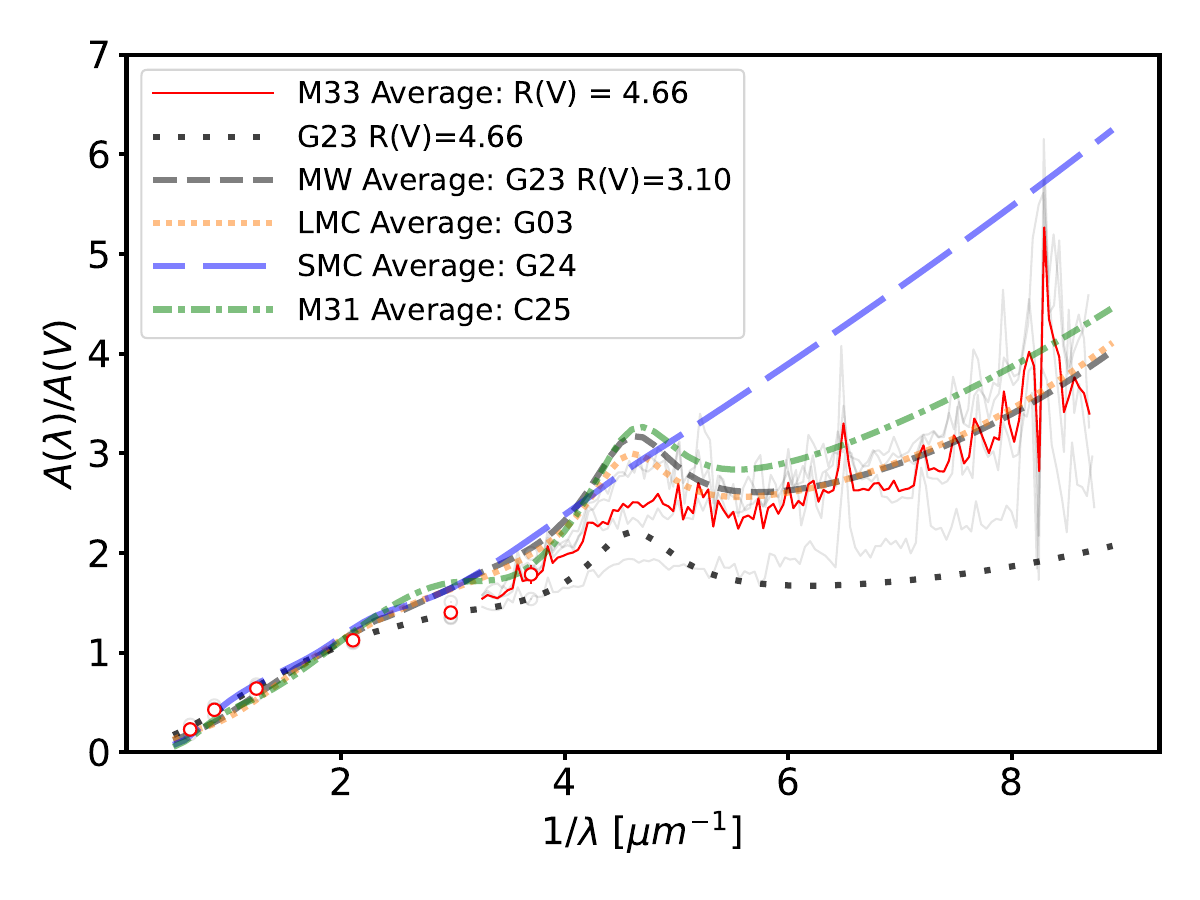}
\caption{The average extinction curve for M33 is shown based on the four measured curves that are shown as faint gray curves.
The G23 MW curve \citep{Gordon23} with the same \rv\ values as measured for the M33 value  is seen to be below the M33 average for all of the UV.
For comparison, all the measured Local Group average curves are plotted \citep{Gordon03, Gordon23, Gordon24, Clayton25}.}
\label{fig_aveext}
\end{figure}

The average of the four M33 sightlines is shown in Fig.~\ref{fig_aveext} along with the averages for the MW, LMC, SMC, and M31 and the predicted curve from the \citet{Gordon23} \rv\ dependent relationship.
The M33 average $R(V) = 4.66$ and FM90 parameters are given in Table~\ref{tab_ext_fm90_params} where FM90 parameters were determined by fitting the \elvebv\ average curve \citep{Gordon24}.

While an average generated from only four sightlines that show large variations is likely to be an approximation to the true average it is still interesting to compute.
The M33 average is lower than all the other averages, most markedly in the blue optical and near-UV.
The average shows a weaker \fbump\ compared to all the other averages except for the SMC.
This is similar to the results found by \citet{Wang22} who also found a weaker \fbump\ than the MW average based on photometry alone.
The average does not follow the MW \rv\ dependent relationship \citep{Gordon23} given the large differences between the $\rv = 4.66$ G23 curve and the average throughout the UV.

\section{Summary}
\label{sec_summary}

The first UV spectroscopic extinction curves for M33 are presented based on new HST/STIS spectroscopic and literature photometric observations towards M33 OB supergiants.
From the observed sample of eight sightlines, it was possible to measure four spectroscopic extinction curves.
These M33 extinction curves are measured using stellar atmosphere models and accounting for the MW foreground extinction.
The dust column parameters \av\ and \rv, gas column \nhi, and detailed FM90 UV extinction shape parameters are given for each curve.

The four extinction curves show strong variations in their overall UV shapes including large variations in \fbump\ strengths and both fairly flat and steep overall UV slopes.
The \rv\ values are within the range of, but higher on average than other Local Group galaxy measurements.
The correlations between the FM90 UV shape parameters fall within the correlations seen for the combination of other Local Group galaxies (MW, LMC, SMC, and M31) and overlap best with the distributions from the LMC and M31.
The correlations between the FM90 UV shape parameters and \nhiav\ fall within the distributions from other Local Group galaxies, again most closely matching the LMC and M31 regions.
The correlation with gas-to-dust ratio is much stronger than the correlation with global metallicity especially given the correlations within an individual galaxy overlap between galaxies with different global metallicities.
The behavior of the correlations with gas-to-dust ratio further strengthens the finding that dust grain formation and destruction through interchange of atoms between the gas and dust phase is happening across the Local Group.
The M33 average extinction curve is seen to be weaker from the blue through the far-UV than other Local Group averages and has a weak \fbump, surpassed only by the SMC.

The code used for the analysis and plots is available\footnote{\url{https://github.com/karllark/hst_m33_ext}}\footnote{\url{https://github.com/karllark/measure_extinction}}\footnote{\url{https://github.com/karllark/extinction_ensemble_props}} \citep{measureextinction, extensembleprops, hstm33ext}.
The STIS data used in this paper can be found in MAST: \dataset[10.17909/t86s-hx97]{http://dx.doi.org/10.17909/t86s-hx97}.
The custom reduced STIS spectra and measured extinction curves are available at \dataset[10.5281/zenodo.16782388]{https://doi.org/10.5281/zenodo.16782388}.
The M33 average extinction curve is available as the G26\_M33Avg average model in the dust\_extinction package\footnote{\url{https://github.com/karllark/dust_extinction}}  \citep{Gordon24JOSS}.

\begin{acknowledgments}

This research is based on observations made with the NASA/ESA Hubble Space Telescope obtained from the Space Telescope Science Institute, which is operated by the Association of Universities for Research in Astronomy, Inc., under NASA contract NAS 5–26555. These observations are associated with GO program 15628. This work was supported by grant, HST-GO-15628.007-A.
PYMJ acknowledges support by the Bulgarian Ministry of Education and Science under the program “Young Scientists and Postdoctoral Scholars”.
MD acknowledges support from the Research Fellowship Program of the European Space Agency (ESA).

This work made use of the following software packages: \texttt{dust\_extinction} \citep{Gordon24JOSS, dustextinction}, \texttt{astropy} \citep{astropy:2013, astropy:2018, astropy:2022}, \texttt{matplotlib} \citep{Hunter:2007}, \texttt{numpy} \citep{numpy}, \texttt{python} \citep{python}, \texttt{scipy} \citep{2020SciPy-NMeth, scipy_15716342}, \texttt{Cython} \citep{cython:2011}, \texttt{emcee} \citep{emcee-Foreman-Mackey-2013, emcee_10996751}, and \texttt{h5py} \citep{collette_python_hdf5_2014, h5py_7560547}.
This research has made use of NASA's Astrophysics Data System.
Software citation information aggregated using \texttt{\href{https://www.tomwagg.com/software-citation-station/}{The Software Citation Station}} \citep{software-citation-station-paper, software-citation-station-zenodo}.

\end{acknowledgments}

\bibliography{dust}
\end{document}